\documentclass[12pt]{article}
\usepackage{graphicx}
\usepackage{amsmath}
\usepackage{lscape}

\newcommand{\bc}{\begin{center}}
\newcommand{\ec}{\end{center}}
\newcommand{\be}{\begin{equation}}
\newcommand{\ee}{\end{equation}}
\newcommand{\bea}{\begin{eqnarray}}
\newcommand{\eea}{\end{eqnarray}}
\newcommand{\beea}{\begin{eqnarray*}}
\newcommand{\eeea}{\end{eqnarray*}}

\newcounter{saveeqn}

\begin{document}

\title{\textbf{\ Evaluation of three-center two-electron repulsion integrals over
Slater orbitals}}
\author{\textbf{Telhat \"OZDO\u{G}AN$\dag$ and Mar\'{\i}a Bel\'{e}n RUIZ} \and 
{\normalsize \noindent Department of Theoretical Chemistry of the
Friedrich-Alexander-University } \\
{\normalsize Erlangen-N\"urnberg, Egerlandstra\ss e 3, D-91058} {\normalsize %
Erlangen, Germany }}
\date{}
\maketitle

\begin{abstract}
\noindent 
The Slater orbitals are the natural basis functions in quantum molecular
calculations. Three-center repulsion Coulomb-exchange integrals over Slater
orbitals are evaluated analytically with  arbitrary orbital exponents,
first for linear conformation of the atomic centers. These integrals have
been expressed as a linear combination of three-center one-electron overlap
integrals, and those have been calculated using auxiliary functions in
terms of one-electron auxiliary integrals. Only one infinite expansion has
been introduced. The resulting integral converges to 20 decimal digits using
about 25-30 terms. Hybrid-exchange three-center repulsion integrals will be
investigated next using this method, as well as triangular conformation of the
centers.
\end{abstract}

\noindent \textbf{keywords}: Slater orbitals; three-center integrals;
two-electron repulsion integrals

\vspace{1cm}

\vspace{5cm}

\noindent $\dag $ Permanent address: Department of Physics, University of
Rize, Turkey. \newline
This work has been supported by DFG and TUBITAK.

\newpage

\section{Introduction}

Slater-type orbitals (STO) \cite{Slater} are the natural basis functions in
quantum molecular calculations. STOs represent well the electron density
near the nucleus (cusp) and far from the nucleus (correct asymptotic
decay). Therefore the STOs resemble the true orbitals. In contrast the
Gaussian type-orbitals (GTO) show a wrong shape near and far of the nucleus
(no cusp). Far of the nucleus the GTOs tend to zero much faster than STOs.

Nevertheless, the use of STOs was hindered in the last four decades by
integration problems and the lack of a product theorem as in the
case of GTOs. Consequently, Slater orbitals were replaced by Gaussian
functions in molecular calculations \cite{Boys}. Despite these
difficulties the research on Slater orbitals has always continued. In 1982,
in a Congress in Tallahassee about Slater orbitals Milan Randic described
the situation: \textit{''Gaussian functions are not the first choice in
theoretical chemistry. They are used (...) primarily because molecular
integrals can be evaluated, not because they posses desirable properties.
Today this may be a valid reason for their use, but tomorrow they may be
thought of as bastard surrogates, which served their purpose in the
transition period, have no longer viable merits and will fail into
oblivion'' }\cite{Randic}. The need of more accurate molecular Configuration
Interaction (CI) calculations and Hartree-Fock (HF) calculations of large
systems has resulted in the use of extremely large basis sets of Gaussian
orbitals and dramatically expensive calculations.

The advantage of using Slater orbitals is that a single basis set would
consist of one function per atom, what would shorten computational times.
Also STOs would provide more accurate description of molecular properties
where the density at the nucleus is important.

One of the main applications of STOs is the accurate
calculation of energy and properties using the CI and
Hylleraas-Configuration Interaction (Hy-CI) methods \cite{SimsH2,Clary}. For
atoms and two-center molecules the neccessary integrals are known. It would
be desirable to extend the application of these methods to three-center
molecules. We plan to test the integrals in the case of the H$_{\mathrm{3}}$
molecule and to use them in CI calculations of the BeH$_{\mathrm{2}}$
molecule.

As it is well-known there are no analytical general methods to calculate the
three- and four-center integrals over Slater orbitals. This problem was soon
recognized in early years (1959) by Mulliken and Roothaan, who called it  
{\it "The bottle-neck of Quantum Chemistry" } \cite{Mulliken}. From the 90s until
today many efforts have been made by several groups to develop efficient
algorithms which have fructified in new computer programs for  
molecules. The bottle-neck of these programs is the calculation of the three-
and four-center repulsion integrals. The approximate methods used are:
one-center expansion \cite
{Coulson,Lowdin,Harris,Guseinov,Smeyers}, translation \cite{Rico,Rico-tcce}, 
Gaussian expansion \cite
{RicoGTO}, Gauss transform \cite{Shavitt,Safouhi,Cesco}, and Fourier transform 
\cite{Filter,Tai,Weniger,Hoggan,Duret,Silverstone} methods. The disadvantages of these methods are
on one hand side the low accuracy achieved, what results in longer
expansions and increased computational times, and on the other  
side the need of numerical integrations. Recently, a special  
issue of a journal \cite{Hoggan-issue} and a book \cite{book} are entirely
dedicated to this topic. Also improvements of the traslation \cite{Rico-tcce}, 
Fourier transform \cite{Silverstone}, and Gauss transform methods \cite{Cesco} 
to calculate the three-center repusion integrals have been recently developed. 

The purpose of this paper is to provide the formulas for the analytical
evaluation of three-center overlap integrals. These integrals are important
because the usual two-electron repulsion integral of the CI method can be
expressed as a linear combination of three-center overlap ones. Therefore
these overlap integrals should be calculated with high accuracy.

In 1936 Hirschfelder {\it et al.} \cite{Hirschfelder,Weygandt} solved some cases of  
two-electron repulsion integrals with equal exponents of $1s$-Slater
orbitals in calculations of the H$_3$ molecule. 
We have extended this method to different
exponents, which is usually the case in quantum mechanical calculations.  
We will treat here the radial part of the integrals. The angular integration
needs to be treated separately because the orientation of the angular functions depends on the geometry.
Also the linear conformation of the centers will be studied first. In Appendix A we give some considerations 
for treating the symmetric triangular case. 

Hybrid-exchange three-center repulsion integrals will be next investigated
using this method. 

\section{Theory}

We consider a linear three-center molecule, see Figure 1. The distances
between the centers may be different and $R=R_{ab}+R_{ac}$. Let $b$ and $c
$ be the focii of an ellipse, the atom $a$ is placed along $R$. The elliptical
coordinates of electron $1$ are defined:

\begin{equation}
\mu_1=\frac{(r_{1b}+r_{1c})}{R},\qquad \nu_1=\frac{(r_{1b}-r_{1c})}{R}%
,\qquad \phi _1=\phi_1
\end{equation}
and the volume element is $R^3/8(\mu^2-\nu^2)d\mu d\nu d\phi .$ The
coordinates of the center $a$ are $\mu_a=1$, $\nu_a=0$ and $\phi_a=0$. The distance
between two points in elliptical coordinates is:

\begin{equation}
r_{1a}^2=\frac{R^2}4\left( \mu _1^2+\mu _a^2+\nu _1^2+\nu _a^2-2-2\mu _1\mu
_a\nu _1\nu _a - [(\mu _1^2-1)(1-\nu _1^2)(\mu
_a^2-1)(1-\nu_a^2)]^{1/2}\cos (\phi _1-\phi _a) \right).
\end{equation}
Substituting the coordinates of $a$ in this equation we obtain:

\begin{equation}
r_{1a}=\frac R2(\mu _1^2+\nu _1^2-1)^{1/2}, 
\end{equation}
\bigskip The three-electron overlap integral over $s$-type Slater orbitals%
\footnote{%
For simplicity only the radial integration will be performed here.} is
defined as:

\begin{equation}
S_{n_an_bn_c}(\zeta_1,\zeta_b,\zeta_c,R)=\int r_a^{n_a-1}e^{-\zeta
_ar_a}r_b^{n_b-1}e^{-\zeta _br_b}r_c^{n_c-1}e^{-\zeta _cr_c}d\tau ,
\end{equation}
where $n_a,n_b,n_c$ are the principal quantum numbers of the three orbitals
located in the centers $a,b,c$, and $\zeta _a,\zeta _b,\zeta _c$ are the
orbital exponents. In the following we will use the notation $r_a =r_{1a}$, $r_b =r_{1b}$, $r_c =r_{1c}$, 
$\mu_1=\mu$ and $\nu_1=\nu$ because the integral depends only on one electron. 
Expressing $r_{b\text{ }}$and $r_c$ in elliptical
coordinates with focii in $b,c$, $r_a$ is already given by Eq. (3):

\begin{equation}
r_b=\frac R2(\mu +\nu ),\qquad r_c=\frac R2(\mu -\nu )
\end{equation}
Substituting Eqs. (3,5) in Eq. (4), writing explicitly the domains of
integration and integrating over $\phi $ we obtain:

\begin{multline}
S_{n_an_bn_c}(\zeta _1,\zeta _b,\zeta _c,R)=\frac{2\pi R^{n_a+n_b+n_c}}{%
2^{n_a+n_b+n_c}}\int_1^\infty d\mu \int_{-1}^1d\nu \left[ (\mu ^2+\nu
^2-1)^{1/2}\right] ^{n_a-1}(\mu +\nu )^{n_b}(\mu -\nu )^{n_c} \\
\times e^{-\alpha \mu -\beta \nu }e^{-\gamma (\mu ^2+\nu ^2-1)^{1/2}}.
\end{multline}
where the term $(\mu ^2-\nu ^2)$ from the volume element has been already
included. $\alpha $, $\beta $ and $\gamma $ are orbital exponents whose 
values are $\alpha =R/2(\zeta _b+\zeta _c)$, $\beta =R/2(\zeta _b-\zeta _c)$
and $\gamma =R/2\zeta _a$. Applying the Binomial Theorem in the form \cite
{Rosen}:

\begin{equation}
(\mu +\nu )^m(\mu -\nu )^n=\sum_{k=0}^{m+n}D_k^{mn}\mu ^{m+n-k}\nu ^k, 
\end{equation}
and defining the coefficients $D_k^{mn}$:

\begin{equation}
D_k^{mn}=\sum_\sigma (-1)^\sigma C_{k-\sigma }^mC_\sigma ^n,\text{ \qquad }%
C_\sigma ^n=\frac{n!}{\sigma !(n-\sigma )!},
\end{equation}
we obtain:

\begin{multline}
S_{n_an_bn_c}(\zeta _1,\zeta _b,\zeta _c,R)=\frac{2\pi R^{n_a+n_b+n_c}}{%
2^{n_a+n_b+n_c}}\sum_{k=0}^{n_b+n_c}D_k^{n_bn_c}\int_1^\infty \mu
^{n_b+n_c-k}e^{-\alpha \mu }d\mu  \\
\times \int_{-1}^1\nu ^k\left[ (\mu ^2+\nu ^2-1)^{1/2}\right]
^{n_a-1}e^{-\beta \nu }e^{-\gamma (\mu ^2+\nu ^2-1)^{1/2}}d\nu .
\end{multline}
Now following Hirschfelder, Eyring and Rosen method \cite{Hirschfelder}, let us make the change of variable:

\begin{equation}
t=(\mu ^2+\nu ^2-1)^{1/2}+\nu .
\end{equation}
The limits are transformed as: $\nu =1\rightarrow $ $t=\mu +1$ and for $\nu
=-1$ $\rightarrow t=\mu -1$. Defining $b=\mu ^2-1$, we have:

\begin{equation}
\nu =\frac 12\left( t-\frac bt\right) ,\text{ \ \qquad }d\nu =\frac 12\left(
1+\frac b{t^2}\right) dt.
\end{equation}
Substituting Eq. (11) in Eq. (10): 

\begin{equation}
(\mu ^2+\nu ^2-1)^{1/2}=\frac 12\left( t+\frac bt\right) ,\text{ }
\end{equation}
and taking into account the transformation of the exponential part of Eq.
(9):

\begin{equation}
e^{-\beta \nu }e^{-\gamma (t-\nu )}\rightarrow e^{-(\beta -\gamma )\nu
}e^{-\gamma t}\rightarrow e^{-\frac{(\beta -\gamma )}2\left( t-\frac
bt\right) }e^{-\gamma t}\rightarrow e^{-\frac{(\beta -\gamma )t}2}e^{-\gamma
t}e^{\frac{\beta -\gamma }2\frac bt}\rightarrow e^{-\frac{(\beta +\gamma )t}%
2}e^{\frac{\beta -\gamma }2\frac bt},
\end{equation}
the three-center overlap integral is:

\begin{multline}
S_{n_an_bn_c}(\zeta _1,\zeta _b,\zeta _c,R)=\frac{2\pi R^{n_a+n_b+n_c}}{%
2^{n_a+n_b+n_c}}\sum_{k=0}^{n_b+n_c}\frac 1{2^k}D_k^{n_bn_c}\int_1^\infty
\mu ^{n_b+n_c-k}e^{-\alpha \mu }d\mu  \\
\times \int_{\mu -1}^{\mu +1}\left( t-\frac bt\right) ^k\left( t+\frac
bt\right) ^{n_a-1}\left( 1+\frac b{t^2}\right) e^{-\frac{(\beta +\gamma )t}%
2}e^{\frac{\beta -\gamma }2\frac bt}dt, 
\end{multline}
where there is an exponential $e^{-\gamma t}$ but also one exponential of
the type $e^{\frac{\beta -\gamma }2\frac bt}$, difficult to integrate.
Therefore we will use the following expansion which converges for a few
terms of the series:

\begin{equation}
e^{\frac{\beta -\gamma }2\frac bt}\simeq \frac 1{N!}\sum_{s=0}^N\frac{N!}{s!}%
\left( \frac{\beta -\gamma }2b\right) ^st^{-s}.
\end{equation}
$N$ is the number of terms of the truncated expansion. Eq. (15) is the only
infinite expansion introduced in the evaluation of the integral. \bigskip
Using the Binomial Theorem:

\begin{equation}
\left( t-\frac bt\right) ^k\left( t+\frac bt\right) ^{n_a-1}\left( 1+\frac
b{t^2}\right) =\frac 1t\left( t+\frac bt\right) ^{n_a}\left( t-\frac
bt\right) ^k=\sum_{u=0}^{n_a+k}D_u^{n_ak}t^{n_a+k-2u-1}b^u,
\end{equation}
we get the integral: 

\begin{eqnarray}
S_{n_an_bn_c}(\zeta _1,\zeta _b,\zeta _c,R) &=&\frac{2\pi R^{n_a+n_b+n_c}}{%
2^{n_a+n_b+n_c}}\frac 1{N!}\sum_{k=0}^{n_b+n_c}\sum_{u=0}^{n_a+k}\sum_{s=0}^N%
\frac{N!}{s!2^k}D_k^{n_bn_c}D_u^{n_ak}\left( \frac{\beta -\gamma }2\right) ^s
\nonumber \\
&&\times \int_1^\infty \mu ^{n_b+n_c-k}(\mu ^2-1)^{u+s}e^{-\alpha \mu }d\mu 
\nonumber \\
&&\times \int_{\mu -1}^{\mu +1}t^{n_a+k-2u-s-1}e^{-\frac{(\beta +\gamma )t}%
2}dt, 
\end{eqnarray}
which can be expressed in terms of auxiliary integrals:  
\begin{eqnarray}
S_{n_an_bn_c}(\zeta _1,\zeta _b,\zeta _c,R) &=&\frac
1{N!}\sum_{k=0}^{n_b+n_c}\sum_{u=0}^{n_a+k}\sum_{s=0}^N\frac{\pi N!}{%
2^{n_a+n_b+n_c+k-1}s!}D_k^{n_bn_c}D_u^{n_ak}\left( \frac{\beta -\gamma }%
2\right) ^s  \nonumber \\
&&\times B_{n_b+n_c-k,s+u;n_a+k-2u-s-1}\left( \alpha ,\frac{(\beta +\gamma )}%
2\right) ,
\end{eqnarray}
where $B_{nm;\pm k}(\alpha ,\beta )$ are auxiliary integrals, which will be
treated in next the Section.

\section{ The auxiliary function $\mathbf{B_{nm;\pm k}}(\mathbf{\alpha
,\beta })$}

We have expressed the three-center overlap integrals as a sum of auxiliary
integrals $B_{nm;\pm k}(\alpha ,\beta )$ which are defined:

\begin{equation}
B_{nm;\pm k}(\alpha ,\beta )=\int_1^\infty \mu ^n(\mu ^2-1)^me^{-\alpha \mu
}d\mu \int_{\mu -1}^{\mu +1}t^{\pm k}e^{-\beta t}dt.
\end{equation}
Two cases are distinguished:  $k$ is positive $t^k$ and the case $t^{-k}$. The
corresponding undefinite integrals are \cite{Tables-ints}:

\begin{equation}
\int t^ke^{-\beta t}dt=-e^{-\beta t}\sum_{h=0}^k\frac{k!}{(h-k)!}\frac{%
t^{k-h}}{\beta ^{h+1}},\text{ \qquad }k\geq 0
\end{equation}
and

\begin{equation}
\int \frac{e^{-\beta t}}{t^k}dt=-e^{-\beta t}\sum_{h=1}^{k-1}\frac{(k-h-1)!}{%
(k-1)!}\frac{(-\beta )^{h-1}}{t^{k-h}}+\frac{(-\beta )^{k-1}}{(k-1)!}%
Ei(-\beta t).\text{\qquad }k>0
\end{equation}
We evaluate now the auxiliary function $B_{nm;+k}(\alpha ,\beta )$ for $%
k\geq 0$ for the actual limits:

\begin{equation}
\int_{\mu -1}^{\mu +1}t^ke^{-\beta t}dt=\sum_{h=0}^k\frac{k!}{(h-k)!}\beta
^{-h-1}\left[ (\mu -1)^{k-h}e^{-\beta (\mu -1)}-(\mu +1)^{k-h}e^{-\beta (\mu
+1)}\right] \text{\qquad }
\end{equation}
substituting in Eq. (19), 
\begin{eqnarray}
B_{nm;+k}(\alpha ,\beta ) &=&\sum_{h=0}^k\frac{k!}{(h-k)!}\beta
^{-h-1}\left( e^\beta \int_1^\infty \mu ^n(\mu +1)^m(\mu
-1)^{m+k-h}e^{-(\alpha +\beta )\mu }d\mu \right.   \nonumber \\
&-&\left. e^{-\beta }\int_1^\infty \mu ^n(\mu +1)^{m+k-h}(\mu
-1)^me^{-(\alpha +\beta )\mu }d\mu \right). 
\end{eqnarray}
The integration over $\mu $ can be written as an one-electron auxiliary
integral which is well-known:

\begin{equation}
B_{nm;+k}(\alpha ,\beta )=\sum_{h=0}^k\frac{k!}{(h-k)!}\beta ^{-h-1}\left(
e^\beta A_{n;m,m+k-h}(\alpha +\beta )-e^{-\beta }A_{n;m+k-h,m}(\alpha +\beta
)\right),
\end{equation}
with 

\begin{equation}
A_{n;mk}(\alpha )=\int_1^\infty \mu ^n(\mu +1)^m(\mu -1)^ke^{-\alpha \mu
}d\mu =\sum_{j=0}^{m+k}D_j^{mk}A_{n+m+k-j}(\alpha ).
\end{equation}

In the case of negative $k$ the calculation is messier:

\begin{multline}
\int_{\mu -1}^{\mu +1}\frac{e^{-\beta t}}{t^k}dt=\frac{(-\beta )^{k-1}}{%
(k-1)!}\left[ Ei(-\beta (\mu +1))-Ei(-\beta (\mu -1))\right]  \\
-\sum_{h=1}^{k-1}\frac{(k-h-1)!}{(k-1)!}(-\beta )^{h-1}\left[ \frac{%
e^{-\beta (\mu +1)}}{(\mu +1)^{k-h}}-\frac{e^{-\beta (\mu -1)}}{(\mu
-1)^{k-h}}\right] ,
\end{multline}
the auxiliary integral is then:

\begin{multline}
B_{nm;-k}(\alpha ,\beta )=\frac{(-\beta )^{k-1}}{(k-1)!}\left[ 
\int_1^\infty \mu ^n(\mu ^2-1)^me^{-\alpha \mu }Ei(-\beta (\mu +1))d\mu \right. \\
-\left. \int_1^\infty \mu ^n(\mu ^2-1)^me^{-\alpha \mu }Ei(-\beta (\mu
-1)d\mu \right]   \\
-\sum_{h=1}^{k-1}\frac{(k-h-1)!}{(k-1)!}(-\beta )^{h-1}\left[ \int_1^\infty
\mu ^n(\mu ^2-1)^me^{-\alpha \mu }\frac{e^{-\beta (\mu +1)}}{(\mu +1)^{k-h}}%
d\mu \right. \\ 
-\left. \int_1^\infty \mu ^n(\mu ^2-1)^me^{-\alpha \mu }\frac{e^{-\beta (\mu
-1)}}{(\mu -1)^{k-h}}d\mu \right] ,
\end{multline}
simplifying:

\begin{multline}
B_{nm;-k}(\alpha ,\beta )=\frac{(-\beta )^{k-1}}{(k-1)!}%
\sum_{i=0}^m(-1)^iC_i^m\left[ \int_1^\infty \mu ^{n+2m-i}e^{-\alpha \mu
}Ei(-\beta (\mu +1))d\mu \right.  \\ 
-\left. \int_1^\infty \mu ^{n+2m-i}e^{-\alpha \mu }Ei(-\beta (\mu -1)d\mu
\right] \\
-\sum_{h=1}^{k-1}\frac{(k-h-1)!}{(k-1)!}(-\beta )^{h-1} 
\left[ e^{-\beta }\int_1^\infty \mu ^n(\mu +1)^{m-k+h}(\mu
-1)^me^{-(\alpha +\beta )\mu }d\mu \right. \\
-\left. e^\beta \int_1^\infty \mu ^n(\mu
+1)^m(\mu -1)^{m-k+h}e^{-(\alpha +\beta )\mu }d\mu \right] .
\end{multline}
Finally:

\begin{multline}
B_{nm;-k}(\alpha ,\beta )=\frac{(-\beta )^{k-1}}{(k-1)!}%
\sum_{i=0}^m(-1)^iC_i^m\left[ T_{n+2m-2i}^{(+)}(\alpha ,\beta
)-T_{n+2m-2i}^{(-)}(\alpha ,\beta )\right]  \\
-\sum_{h=1}^{k-1}\frac{(k-h-1)!}{(k-1)!}(-\beta )^{h-1}\left[ e^{-\beta
}A_{n;m-k+h,m}(\alpha +\beta )-e^\beta A_{n;m,m-k+h}(\alpha +\beta )\right] 
\end{multline}
the auxiliary integral for negative values of $k$ is a sum of one-electron
auxiliary integrals $A_{n;mk}(\alpha +\beta )$ and a new kind of auxiliary
integrals, which will be evaluated in the next Section. 

\subsubsection*{3.1 Calculation of $\mathbf{T_n^{\mathbf{(+)}}({\bf \alpha ,\beta}
)}$ and $\mathbf{T_n^{(+)}}({\bf \alpha ,\beta })$}

The auxilary functions $T_{n+2m-2i}^{(+)}(\alpha ,\beta )$ and $%
T_{n+2m-2i}^{(-)}(\alpha ,\beta )$ need to be calculated. They are defined
as:

\begin{eqnarray}
T_n^{(+)}(\alpha ,\beta ) &=&\int_1^\infty \mu ^ne^{-\alpha \mu }Ei(-\beta
(\mu +1)d\mu , \\
T_n^{(-)}(\alpha ,\beta ) &=&\int_1^\infty \mu ^ne^{-\alpha \mu }Ei(-\beta
(\mu -1)d\mu .  \nonumber
\end{eqnarray}
The auxiliary functions \bigskip $T_n^{(+)}(\alpha ,\beta )$ can be
calculated using recursion relations. The first term is:

\begin{equation}
T_0^{(+)}(\alpha ,\beta )=\int_1^\infty e^{-\alpha \mu }Ei(-\beta (\mu
+1)d\mu .
\end{equation}
The higher terms can be calculated using the differential equation:

\begin{equation}
T_n^{(+)}(\alpha ,\beta )=(-1)^n\frac{d^n}{d\alpha ^n}T_0^{(+)}(\alpha
,\beta ).
\end{equation}
Let us first solve $T_0^{(+)}(\alpha ,\beta )$, making the change of
variables:

\begin{equation}
\beta (\mu +1)=y,\text{ \qquad }\beta d\mu =dy,\qquad \mu =\frac y\beta -1,
\end{equation}
the integration limits of the domains are transformed $\mu =1\rightarrow
y\Rightarrow 2\beta $ and $\mu \rightarrow \infty \Rightarrow y\rightarrow
\infty $. To calculate $T_0^{(+)}(\alpha ,\beta )$ one has to solve the
integral: 

\begin{equation}
T_0^{(+)}(\alpha ,\beta )=\int_{2\beta }^\infty \frac 1\beta e^{-\alpha
\left( \frac y\beta -1\right) }Ei(-y)dy.
\end{equation}
Using the integration formula: 
\be
\int_c^\infty \frac 1\beta e^{-\gamma y}Ei(-y)dy=\frac 1\gamma \left[
Ei(-c)e^{-\gamma c}-cEi(-(\gamma +1))\right] ,
\ee
we have

\begin{equation}
T_0^{(+)}(\alpha ,\beta )=\frac{e^\alpha }\beta \frac 1{(\alpha /\beta
)}\left[ Ei(-2\beta )e^{-\frac \alpha \beta 2\beta }-Ei\left( -\left( \frac
\alpha \beta +1\right) 2\beta \right) \right] ,
\end{equation}
and simplifying:\bigskip 
\begin{equation}
T_0^{(+)}(\alpha ,\beta )=\frac{e^{-\alpha }}\alpha Ei(-2\beta )-\frac{%
e^\alpha }\alpha Ei(-2(\alpha +\beta )).
\end{equation}

The $n$th-term is: 

\begin{equation}
T_n^{(+)}(\alpha ,\beta )=(-1)^n\frac{d^n}{d\alpha ^n}\left( \frac{%
e^{-\alpha }}\alpha Ei(-2\beta )-\frac{e^\alpha }\alpha Ei(-2(\alpha +\beta
))\right) .
\end{equation}
Let us use the Leibnitz formula:

\begin{equation}
\frac{d^n}{dx^n}\left( f(x)g(x)\right) =\sum_{m=0}^nC_m^n\frac{d^{n-m}}{%
dx^{n-m}}f(x)\frac{d^m}{dx^m}g(x)
\end{equation}
calculating $\frac{d^n}{dx^n}\frac{e^{-\alpha }}\alpha Ei(-2\beta )$ and $%
\frac{d^n}{dx^n}\frac{e^\alpha }\alpha Ei(-2(\alpha +\beta ))$:

\begin{equation}
(-1)^n\frac{d^n}{d\alpha ^n}\frac{e^{-\alpha }}\alpha Ei(-2\beta )=(-1)^n%
\frac{d^n}{d\alpha ^n}\left( A_0(\alpha )Ei(-2\beta )\right) =A_n(\alpha
)Ei(-2\beta )
\end{equation}
and

\begin{multline}
(-1)^n\frac{d^n}{d\alpha ^n}\frac{e^\alpha }\alpha Ei(-2(\alpha +\beta ))
=(-1)^n\frac{d^n}{d\alpha ^n}\left( -A_0(-\alpha )Ei(-2(\alpha +\beta
))\right) \\                   
=(-1)^{n+1}\sum_{m=0}^nC_m^n\frac{d^{n-m}}{d\alpha ^{n-m}}A_0(-\alpha )%
\frac{d^m}{d\alpha ^m}Ei(-2(\alpha +\beta ))  \\                  
=-A_n(-\alpha )Ei(-2(\alpha +\beta
))+\sum_{m=1}^n(-1)^{n-m}C_m^nA_{n-m}(-\alpha )Ei(-2(\alpha +\beta )). 
\end{multline}
We have used the following definitions of the exponential integral: 

\begin{equation}
Ei(-z)=\int_{-\infty }^z\frac{e^{-t}}tdt=-\int_z^\infty \frac{e^{-t}}tdt
\end{equation}

\begin{equation}
E_1(-z)=\int_z^\infty \frac{e^{-t}}tdt=-Ei(-z)\equiv Ei(1,z).
\end{equation}
The derivatives of the exponential integral are

\begin{equation}
\frac d{d\alpha }Ei(-2(\alpha +\beta ))=(-1)(-2)\int_1^\infty e^{-2(\alpha
+\beta )t}dt=2A_0(2(\alpha +\beta )),
\end{equation}

\begin{equation}
\frac{d^m}{d\alpha ^m}Ei(-2(\alpha +\beta ))=\frac{d^{m-1}}{d\alpha ^{m-1}}%
2A_0(2(\alpha +\beta ))=(-1)^{m-1}2^mA_{m-1}(2(\alpha +\beta ))
\end{equation}

\begin{multline}
(-1)^n\frac{d^n}{d\alpha ^n}\left( \frac{e^\alpha }\alpha Ei(-2(\alpha
+\beta ))\right) =-(-1)^nA_n(-\alpha )Ei(-2(\alpha +\beta )) \\
+\sum_{m=1}^n(-1)^{n-m}2^mC_m^nA_{n-m}(-\alpha )(-1)^{m-1}A_{m-1}(2(\alpha
+\beta )).
\end{multline}

Finally we have: 

\begin{eqnarray}
T_n^{(+)}(\alpha ,\beta ) &=&A_n(-\alpha )Ei(-2\beta )+(-1)^nA_n(-\alpha
)Ei(-2(\alpha +\beta ))  \nonumber \\
&&-\sum_{m=1}^n(-1)^{n-m}2^mC_m^nA_{n-m}(-\alpha )(-1)^{m-1}A_{m-1}(2(\alpha
+\beta ))
\end{eqnarray}

\bigskip To calculate the auxiliary functions $T_n^{(-)}(\alpha ,\beta )$
defined as:

\begin{equation}
T_n^{(-)}(\alpha ,\beta )=\int_1^\infty \mu ^ne^{-\alpha \mu }Ei(-\beta (\mu
-1)d\mu
\end{equation}
\bigskip we use also recursion relations. Let us calculate also $%
T_0^{(-)}(\alpha ,\beta )$ defined as: 

\begin{equation}
T_0^{(-)}(\alpha ,\beta )=\int_1^\infty e^{-\alpha \mu }Ei(-\beta (\mu
-1)d\mu .
\end{equation}
\bigskip The following change of variable is done:

\begin{eqnarray}
u &=&Ei(-\beta (\mu -1)),\qquad \Rightarrow du=\frac{e^{-\beta (\mu -1)}}{%
(\mu -1)} \\
dv &=&e^{-\alpha \mu },\qquad \Rightarrow v=-\frac{e^{-\alpha \mu }}\alpha .  
\nonumber
\end{eqnarray}
The integral $T_0^{(-)}(\alpha ,\beta )$ is the limit:  

\bigskip 
\begin{equation}
T_0^{(-)}(\alpha ,\beta )=\lim_{\gamma \rightarrow 1}\int_\gamma ^\infty
e^{-\alpha \mu }Ei(-\beta (\mu -1)d\mu ,
\end{equation}
integrating by parts:

\begin{equation}
T_0^{(-)}(\alpha ,\beta )=\lim_{\gamma \rightarrow 1}\left( -\frac 1\alpha
e^{-\alpha \mu }Ei(-\beta (\mu -1)+\frac 1\alpha \int_\gamma ^\infty
e^{-\alpha \mu }\frac{e^{-\beta (\mu -1)}}{(\mu -1)}d\mu \right)
\end{equation}
making again a change of variables, in this case: $\mu -1=u\Rightarrow d\mu
=du$ and $\mu =\gamma \Rightarrow u=\gamma -1$ with $\mu \rightarrow \infty
\Rightarrow u\rightarrow \infty $, we have: 

\begin{multline}
T_0^{(-)}(\alpha ,\beta )=\lim_{\gamma \rightarrow 1}\left( -\frac 1\alpha
e^{-\alpha \mu }Ei(-\beta (\mu -1)+\frac{e^{-\alpha }}\alpha \int_{\gamma
-1}^\infty \frac{e^{-(\alpha +\beta )u}}udu\right)  \\
=\frac{e^{-\alpha }}\alpha \lim_{\gamma \rightarrow 1}\left( Ei(-\beta (\mu
-1)-Ei(-(\alpha +\beta )(\mu -1)\right)  \\
=\lim_{\gamma \rightarrow 1}\left( -\int_{\beta (\gamma -1)}^\infty \frac{%
e^{-t}}tdt+\int_{(\alpha +\beta )(\gamma -1)}^\infty \frac{e^{-t}}tdt\right)
,
\end{multline}
using the identity

\begin{equation}
-\int_a^\infty \frac{e^{-t}}tdt=\int_a^1\left( \frac{1-e^{-t}}t\right)
dt-\int_1^\infty \frac{e^{-t}}tdt-\int_a^1\frac{dt}t
\end{equation}
and the definition of the Euler constant $C$:

\begin{equation}
C=\int_a^1\left( \frac{1-e^{-t}}t\right) dt-\int_1^\infty \frac{e^{-t}}%
tdt=0.5772156649015. 
\end{equation}
we get:          

\begin{multline}
T_0^{(-)}(\alpha ,\beta )=\frac{e^{-\alpha }}\alpha \lim_{\gamma \rightarrow
1}\left\{ \int_{\beta (\gamma -1)}^1\frac{1-e^{-t}}tdt-\int_1^\infty \frac{%
e^{-t}}tdt-\int_{\beta (\gamma -1)}^1\frac{dt}t-\int_{(\alpha +\beta
)(\gamma -1)}^\infty \frac{1-e^{-t}}tdt\right.  \\
+\left. \int_1^\infty \frac{e^{-t}}tdt+\int_{(\alpha +\beta )(\gamma -1)}^1%
\frac{e^{-t}}tdt\right\} 
\end{multline}

\begin{equation}
T_0^{(-)}(\alpha ,\beta )=\frac{e^{-\alpha }}\alpha \lim_{\gamma \rightarrow
1}\left( C-C-\left. \ln t\right| _{\beta (\gamma -1)}^1+\left. \ln t\right|
_{(\alpha +\beta )(\gamma -1)}^1\right) 
\end{equation}
simplifying: 

\begin{multline}
T_0^{(-)}(\alpha ,\beta )=\frac{e^{-\alpha }}\alpha \lim_{\gamma \rightarrow
1}\left( \ln [\beta (\gamma -1)]-\ln [(\alpha +\beta )(\gamma -1)]\right)  \\
=\frac{e^{-\alpha }}\alpha \lim_{\gamma \rightarrow 1}\left( \ln \beta +\ln
(\gamma -1)-\ln (\alpha +\beta )-\ln (\gamma -1)\right) 
\end{multline}
we finally get for $T_0^{(-)}(\alpha ,\beta )$ a simple expression: 

\be 
T_0^{(-)}(\alpha ,\beta )=\frac{e^{-\alpha }}\alpha \ln \left( \frac \beta
{\alpha +\beta }\right) 
\ee 

Now we consider the $n$th-term: 
\begin{eqnarray}
T_n^{(-)}(\alpha ,\beta ) &=&(-1)^n\frac{d^n}{d\alpha ^n}T_0^{(-)}(\alpha
,\beta )=(-1)^n\frac{d^n}{d\alpha ^n}\frac{e^{-\alpha }}\alpha \ln \left(
\frac \beta {\alpha +\beta }\right)   \nonumber \\
&=&(-1)^n\frac{d^n}{d\alpha ^n}\left[ A_0(\alpha )\ln \left( \frac \beta
{\alpha +\beta }\right) \right]   \nonumber \\
&=&(-1)^n\sum_{m=0}^nC_m^n\frac{d^{n-m}}{d\alpha ^{n-m}}A_0(\alpha )\frac{d^m%
}{d\alpha ^m}\ln \left( \frac \beta {\alpha +\beta }\right) 
\end{eqnarray}
obtaining finally: \bigskip 
\begin{equation}
T_n^{(-)}(\alpha ,\beta )=A_n(\alpha )\ln \left( \frac \beta {\alpha +\beta
}\right) +\sum_{m=1}^nC_m^n\frac{(m-1)!}{(\alpha +\beta )^m}A_{n-m}(\alpha )
\end{equation}
where  $A_n(\alpha )$ is the Mulliken integral:

\begin{equation}
A_n(\alpha )=\int_1^\infty t^ne^{-\alpha t}=\frac{n!e^{-\alpha }}{\alpha
^{n+1}}\sum_{k=0}^n\frac{\alpha ^k}{k!}.
\end{equation}

In Table 1, the pattern of convergence of the three-center overlap integral
is shown. To obtain a convergence of 15 decimal digits 19 terms of the
expansion are needed. To obtain 30 decimal digits accuracy, about 30
terms of the expansion are required. These values have been calculated with 
Maple \cite{Maple}. This algorithm will be programmed in a Fortran source code  
and the efficiency of the calculation of the integral can be tested.
Further, this integral will be tested by comparing the two-electron
three-center integrals with the values obtained by the SMILES computer code 
\cite{Rico-prog} version which uses the Gaussian expansion method. 

In Table 2, a number of three-center one-electron integrals for different
exponents, quantum numbers and interatomic distances is presented, with respect to  
the number of terms neccessary in the expansion. 

\subsubsection*{3.2 Summary}

The three-electron overlap pone-electron integral is: 

\begin{eqnarray}
S_{n_an_bn_c}(\zeta _1,\zeta _b,\zeta _c,R) &=&\frac
1{N!}\sum_{k=0}^{n_b+n_c}\sum_{u=0}^{n_a+k}\sum_{s=0}^N\frac{\pi N!}{%
2^{n_a+n_b+n_c+k-1}s!}D_k^{n_bn_c}D_u^{n_ak}\left( \frac{\beta -\gamma }%
2\right) ^s  \nonumber \\
&&\times B_{n_b+n_c-k,s+u;n_a+k-2u-s-1}\left( \alpha ,\frac{(\beta +\gamma )}%
2\right) ,
\end{eqnarray}
The auxiliary integrals for postive and negative $k$:

\begin{equation}
B_{nm;+k}(\alpha ,\beta )=\sum_{h=0}^k\frac{k!}{(h-k)!}\beta ^{-h-1}\left(
e^\beta A_{n;m,m+k-h}(\alpha +\beta )-e^{-\beta }A_{n;m+k-h,m}(\alpha +\beta
)\right)
\end{equation}

\begin{multline}
B_{nm;-k}(\alpha ,\beta )=\frac{(-\beta )^{k-1}}{(k-1)!}%
\sum_{i=0}^m(-1)^iC_i^m\left[ T_{n+2m-2i}^{(+)}(\alpha ,\beta
)-T_{n+2m-2i}^{(-)}(\alpha ,\beta )\right]  \\
-\sum_{h=1}^{k-1}\frac{(k-h-1)!}{(k-1)!}(-\beta )^{h-1}\left[ e^{-\beta
}A_{n;m-k+h,m}(\alpha +\beta )-e^\beta A_{n;m,m-k+h}(\alpha +\beta )\right] 
\end{multline}
The auxiliary functions: 

\begin{eqnarray}
T_n^{(+)}(\alpha ,\beta ) &=&A_n(-\alpha )Ei(-2\beta )+(-1)^nA_n(-\alpha
)Ei(-2(\alpha +\beta ))  \nonumber \\
&&-\sum_{m=1}^n(-1)^{n-m}2^mC_m^nA_{n-m}(-\alpha )(-1)^{m-1}A_{m-1}(2(\alpha
+\beta ))
\end{eqnarray}

\begin{equation}
T_n^{(-)}(\alpha ,\beta )=A_n(\alpha )\ln \left( \frac \beta {\alpha +\beta
}\right) +\sum_{m=1}^nC_m^n\frac{(m-1)!}{(\alpha +\beta )^m}A_{n-m}(\alpha )
\end{equation}
The one-electron auxiliary integrals:

\begin{equation}
A_n(\alpha )=\int_1^\infty t^ne^{-\alpha t}=\frac{n!e^{-\alpha }}{\alpha
^{n+1}}\sum_{k=0}^n\frac{\alpha ^k}{k!},
\end{equation}

\begin{equation}
A_{n;mk}(\alpha )=\sum_{j=0}^{m+k}D_j^{mk}A_{n+m+k-j}(\alpha ).
\end{equation}

\section{Two-electron three-center integrals}

There are several kinds of two-electron integrals where three centers are
involved, see Figure 1: Coulomb-exchange (the first two) and hybrid-exchange
(the last two):

\begin{eqnarray}
&&[aa,bc]=\int \int \chi _a(1)\chi _a(1)\frac 1{r_{12}}\chi _b(2)\chi
_c(2)d\tau _1d\tau _2,  \nonumber \\
&&[bb,ac]=\int \int \chi _b(1)\chi _b(1)\frac 1{r_{12}}\chi _a(2)\chi
_c(2)d\tau _1d\tau _2,  \nonumber \\
&&[ab,ac]=\int \int \chi _a(1)\chi _b(1)\frac 1{r_{12}}\chi _a(2)\chi
_c(2)d\tau _1d\tau _2,  \nonumber \\
&&[ab,bc]=\int \int \chi _a(1)\chi _b(1)\frac 1{r_{12}}\chi _b(2)\chi
_c(2)d\tau _1d\tau _2.
\end{eqnarray}
The Coulomb exchange integrals can be evaluated by integration over the coordinates of electron $1$ \cite{Peuker}, 
obtaining a linear combination of three-center overlap integrals calculated in this paper:

In general a two-electron integral can be written:

\begin{equation}
[aa,bc]= \int_0^{\infty} dr_2 \int_{r_2}^{\infty} \chi_a (1) \chi_a(1) \hat
g_{12} \chi_b (2) \chi_c (2) d\tau_1 d\tau_2,
\end{equation}
where $\hat g_{12}$ is an operator of $r_1$ and $r_2$. For $L=0$ the
operator is $\hat g_{12}=g_{12}$ where $g_{12}$ means the largest of $r_1$
and $r_2$. There are two integration domains: $r_1 < r_2 $ and $r_1 > r_2$.
One may refer the interelectronic distance $r_{12}$ to center $a$ and expand 
it in terms of the variables $r_{1a}$ and $r_{2a}$:   

\begin{multline}
[aa,bc]= \int_0^{\infty} dr_1 \int_{r_{1a}}^{\infty} \chi_a (1) \chi_a(1) \frac{%
1}{r_2} \chi_b (2) \chi_c (2) dr_{2a} + \int_0^{\infty} dr_{2a}
\int_{r_{2a}}^{\infty} \chi_a (1) \chi_a(1) \frac{1}{r_{1a}} \chi_b (2) \chi_c (2)
dr_{1a} \\
= \int_0^{\infty} dr_{2a} \int_0^{r_{2a}} \chi_a (1) \chi_a(1) \frac{1}{r_{2a}}
\chi_b (2) \chi_c (2) dr_{1a} + \int_0^{\infty} dr_{2a} \int_{r_{2a}}^{\infty} \chi_a
(1) \chi_a(1) \frac{1}{r_{1a}} \chi_b (2) \chi_c (2) dr_{1a}. 
\end{multline}
As these orbitals are located at the same center, we can integrate  
over the coordinates of one-electron \cite{Peuker}:

\begin{equation}
A_n(x,\alpha )=\int_x^{\infty} r_{1a}^ne^{-\alpha r_{1a}}dr_{1a}=e^{-\alpha
x} \sum_{k=0}^n\frac{n!}{(n-k)!}\frac{x^{n-k}}{\alpha^{k+1}}, 
\end{equation}
\begin{equation}
U_n(x,\alpha )=\int_0^x r_{1a}^ne^{-\alpha
r_{1a}}dr_{1a}=A_n(\alpha)-A_n(x,\alpha) =\frac{n!}{\alpha^{n+1}}-
e^{-\alpha x} \sum_{k=0}^n\frac{n!}{(n-k)!}\frac{x^{n-k}}{\alpha^{k+1}}.
\end{equation}
The Coulomb-exchange integrals of Eq. (70) can be evaluated integrating over the coordinates of electron $1$,
writting them explicitly: $n_{1a}=n_1+n_1^{\prime}-1$ and $\zeta=\zeta_1 +
\zeta_1^{\prime}$:

\begin{eqnarray}
&&[aa,bc] = 4\pi \int r_{2b}^{n_b-1} e^{-\zeta_b r_2b} r_{2c}^{n_c-1}
e^{-\zeta_c r_2c} d\tau_2 \int r_{1a}^{n_a-1} e^{-\zeta_a r_{1a}} \frac{1}{%
r_{12}} r_{1a}^2 dr_{1a},  \nonumber \\
&&[bb,ac] = 4\pi\int r_{2a}^{n_a-1} e^{-\zeta_a r_2a} r_{2c}^{n_c-1}
e^{-\zeta_c r_2c} d\tau_2 \int r_{1b}^{n_b-1} e^{-\zeta_b r_{1b}} \frac{1}{%
r_{12}} r_{1b}^2 dr_{1b},
\end{eqnarray}
using Eqs. (73) and (74) we obtain

\begin{equation}
[aa,bc] = 4\pi \int_0^{\infty} r_{2b}^{n_b-1} e^{-\zeta_b r_2b}
r_{2c}^{n_c-1} e^{-\zeta_c r_2c} d\tau_2 \left\{\frac{1}{r_{2a}}
\int_0^{r_{2a}} r_{1a}^{n_a+1} e^{-\zeta_a r_{1a}} dr_{1a}
+\int_{r_{2a}}^{\infty} r_{1a}^{n_a} e^{-\zeta_a r_{1a}} dr_{1a} \right\}, 
\end{equation}

\begin{multline}
[aa,bc] = 4\pi \int_0^{\infty} r_{2b}^{n_b-1} e^{-\zeta_b r_2b}
r_{2c}^{n_c-1} e^{-\zeta_c r_2c} d\tau_2 \left\{\frac{1}{r_{2a}} \frac{%
(n_a+1)!}{\zeta_a^{n_a+2}} -\frac{e^{-\zeta_a r_{2a}}}{r_{2a}} \frac{(n_a+1)!%
}{\zeta_a^{n_a+2}} \right. \\
\left.- \sum_{k=0}^{n_a+1}\frac{(n_a+1)!}{(n_a+1-k)!\zeta_a^{k+1}}%
r_{2a}^{n_a-k}e^{-\zeta_a r_{2a}} +\sum_{k=0}^{n_a}\frac{(n_a)!}{%
(n_a-k)!\zeta_a^{k+1}}r_{2a}^{n_a-k}e^{-\zeta_a r_{2a}} \right\},
\end{multline}

\begin{multline}
[aa,bc] = 4\pi \int_0^{\infty} r_{2b}^{n_b-1} e^{-\zeta_b r_2b}
r_{2c}^{n_c-1} e^{-\zeta_c r_2c} d\tau_2 \left\{\frac{1}{r_{2a}} \frac{%
(n_a+1)!}{\zeta_a^{n_a+2}} \left(1-e^{-\zeta_a r_{2a}}\right) \right. \\
\left. +\sum_{k=0}^{n_a}\frac{k-1}{(n_a+1-k)}\frac{(n_a)!}{%
(n_a-k)!\zeta_a^{k+1}}r_{2a}^{n_a-k}e^{-\zeta_a r_{2a}} \right\}. 
\end{multline}

Finally the Coulomb-exchange reulsion integral is expressed as a linear combination 
of three-center overlap integrals defined in Eq. (4): 
\begin{multline}
[aa,bc] = 4\pi \frac{(n_a+1)!}{\zeta_a^{n_a+2}} \left(
S_{0,n_b,n_c}(0,\zeta_b,\zeta_c,R) -
S_{0,n_b,n_c}(\zeta_a,\zeta_b,\zeta_c,R)\right) \\
+4\pi \sum_{k=0}^{n_a}\frac{k-1}{(n_a+1-k)}\frac{n_a!}{(n_a-k)!%
\zeta_a^{k+1}} S_{n_a-k+1,n_b,n_c}(\zeta_a,\zeta_b,\zeta_c,R),
\end{multline}
where the integral $S_{0,n_b,n_c}(0,\zeta_b,\zeta_c,R)$ is the three-center 
nuclear attraction integral defined in Ref. \cite{Peuker-tcna}:  
\begin{equation}
S_{0,n_b,n_c}(0,\zeta_b,\zeta_c,R)=K_{na,nbnc} =\int \frac{\chi_b \chi_c}{r_a%
} d\tau
\end{equation}
$S_{0,n_b,n_c}(\zeta_a,\zeta_b,\zeta_c,R))$ should be analyzed. Anagously
for $[bb,ac]$ we obtain:

\begin{multline}
[bb,ac] = 4\pi \frac{(n_b+1)!}{\zeta_b^{n_b+2}} \left(
S_{0,n_a,n_c}(0,\zeta_a,\zeta_c,R) -
S_{0,n_a,n_c}(\zeta_b,\zeta_a,\zeta_c,R)\right) \\
+4\pi \sum_{k=0}^{n_b}\frac{k-1}{(n_b+1-k)}\frac{n_b!}{(n_b-k)!%
\zeta_b^{k+1}} S_{n_b-k+1,n_a,n_c}(\zeta_b,\zeta_a,\zeta_c,R)
\end{multline}
$S_{0,n_a,n_c}(0,\zeta_a,\zeta_c,R)$ are expressed as: 
\begin{equation}
S_{0,n_a,n_c}(0,\zeta_a,\zeta_c,R)=K_{nb,nanc} =\int \frac{\chi_a \chi_c}{r_b%
} d\tau
\end{equation}
The integral $S_{0,n_a,n_c}(\zeta_b,%
\zeta_a,\zeta_c,R)$ is an especial case of overlap integral with $n_a=-1$ and 
should be treated separately, as Eqs. (9-14) are derived for $n_a \ge 1$.  

Next we will calculate some two-electron integrals and compare with
the literature values and values obtained with a computer program by Rico {\it et al.}  
\cite{Rico-prog} using the Gaussian expansion method.

\section*{Conclusions}

In this paper three-center overlap integrals are calculated, which are used to 
calculate  
the two-center three-center repulsion integrals. The integrals consist of an
expansion which converges relatively fast.  

\section*{Acknowledments}

One of us (T. \"{O}.) would like to thank the Deutsche
Forschungsgemeinschaft and TUBITAK for a grant to visit Erlangen for three
months. The authors are very indepted to Prof. Peter Otto for encouraging
this project.

\section*{Appendix A}

\setcounter{equation}{0}

\renewcommand{\theequation}{A.\arabic{equation}}

In Figure 2 the symmetric triangular conformation of the centers is shown. The interelectronic distances 
are equal $R=R_{ab}=R_{ac}$. The elliptical coordinates of the center $a$ with respect to the focii $b,c$ 
are fixed values: 
\be
\mu_a=\frac{R_{ab}+R_{ac}}{R}=2, \qquad \nu_a=0,
\ee
and the relation:
\be
\cos (\phi_1 - \phi_a )= \cos \phi_1 \cos \phi_a + \sin \phi_1 \sin \phi_a
\ee
independently of the choose of the angle $\phi_a$, in integrations over $s$-type Slater the integrals  
$\int_0^{2\pi}\cos \phi_1d\phi_1 =0$ and $\int_0^{2\pi}\sin \phi_1d\phi_1 =0$ vanish. Then we have: 
\begin{equation}
r_{1a}^2=\frac{R^2}4 ( \mu _1^2+\nu _1^2+2 )             
\end{equation}
and therefore 
\begin{equation}
r_{1a}=\frac R2(\mu _1^2+\nu _1^2+2)^{1/2},
\end{equation}
accordingly the substitution in this case is: 
\begin{equation}
t=(\mu ^2+\nu ^2+2)^{1/2}+\nu .
\end{equation}
which is similar to the one used in Section 1, and therefore the steps of integration described 
in previous Sections can be used. 
The triangular conformation is very important because it is the ground state conformation of the H$_3$ molecule 
and many molecular systems.

\newpage

\begin{table}[tbp]
\begin{center}
\textbf{Table 1}: Convergence pattern of the three-center overlap integral for
the orbital parameters $n_a=1$, $n_b=2$, $n_c=1$ and $\zeta_a=1.6$, $\zeta_b=1.4$, $\zeta_c=1.2$; $%
R=1.4$ a.u.\\[0pt]
\begin{tabular}{rl}
\hline\hline
1 & -0.11416 \\ 
2 & -0.12293 3 \\ 
3 & -0.12160 22 \\ 
4 & -0.12219 350 \\ 
5 & -0.12207 7743 \\ 
6 & -0.12210 04107 \\ 
7 & -0.12209 67757 3 \\ 
8 & -0.12209 73204 83 \\ 
9 & -0.12209 72448 624 \\ 
10 & -0.12209 72547 7065 \\ 
11 & -0.12209 72535 25580 \\ 
12 & -0.12209 72536 75642 0 \\ 
13 & -0.12209 72536 58049 61 \\ 
14 & -0.12209 72536 60052 409 \\ 
15 & -0.12209 72536 59828 9497 \\ 
16 & -0.12209 72536 59853 35836 \\ 
17 & -0.12209 72536 59850 73400 8 \\ 
18 & -0.12209 72536 59851 01158 63 \\ 
19 & -0.12209 72536 59850 98261 802 \\ 
20 & -0.12209 72536 59850 98559 9711 \\ 
21 & -0.12209 72536 59850 98529 65861 \\ 
22 & -0.12209 72536 59850 98532 70074 8 \\ 
23 & -0.12209 72536 59850 98532 39925 80 \\ 
24 & -0.12209 72536 59850 98532 42873 615 \\ 
25 & -0.12209 72536 59850 98532 42589 4768 \\ 
26 & -0.12209 72536 59850 98532 42616 42719 \\ 
27 & -0.12209 72536 59850 98532 42613 91777 \\ 
28 & -0.12209 72536 59850 98532 42614 14624 \\ 
29 & -0.12209 72536 59850 98532 42614 12602 \\ 
30 & -0.12209 72536 59850 98532 42614 12774 \\ \hline\hline
\end{tabular}
\end{center}
\end{table}

\begin{table}[tbp]
\begin{center}
\textbf{Table 2}: Three-center overlap integrals. $N$ is the   
number of terms of the expansion needed to calculate accurately the printed number of 
decimal digits. \\[0pt]
\begin{tabular}{ccccccclr}
\hline\hline
$n_a$ & $n_b$ & $n_c$ & $\zeta_a$ & $\zeta_b$ & $\zeta_c$ & $R$ & $%
S_{n_an_bn_c}(\zeta _1,\zeta _b,\zeta _c,R)$ & N \\ \hline               
1 & 1 & 1 & 1.3 & 1.3 & 1.3 & 1.0 &    -0.10852 96351 53609 38988 & 25 \\
1 & 1 & 1 & 1.1 & 1.3 & 1.5 & 1.0 &    -0.41057 77481 70340 63986 & 26 \\
1 & 2 & 2 & 2.0 & 2.0 & 2.0 & 2.0 & $\;$0.00409 87551 44074 883  & 30 \\
2 & 2 & 2 & 1.6 & 1.4 & 1.4 & 2.5 & $\;$0.08943 10170 26917 0505 & 28 \\ 
4 & 3 & 2 & 2.6 & 2.4 & 1.6 & 3.0 & $\;$0.00362 95811 14392 3873 & 25 \\ 
\hline\hline
\end{tabular}
\end{center}
\end{table}

\newpage

\begin{figure}[]
\includegraphics[scale=0.90]{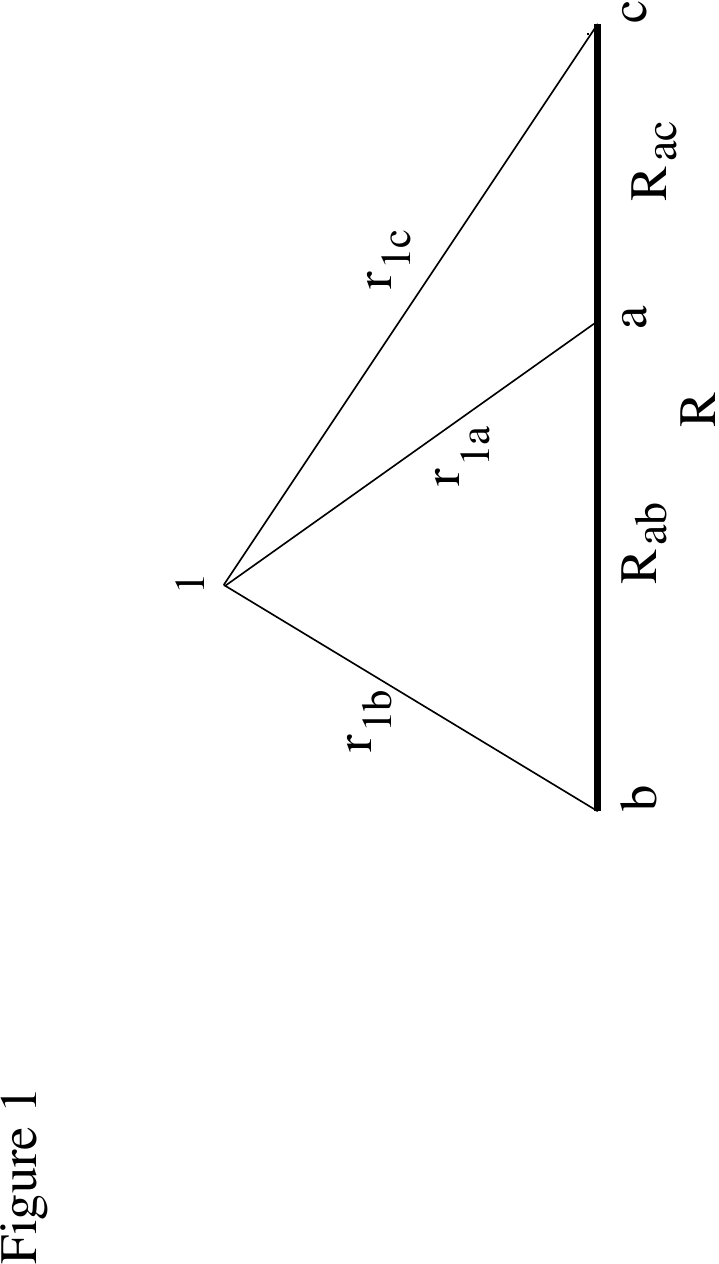}
\end{figure}

\begin{figure}[]
\includegraphics[scale=0.90]{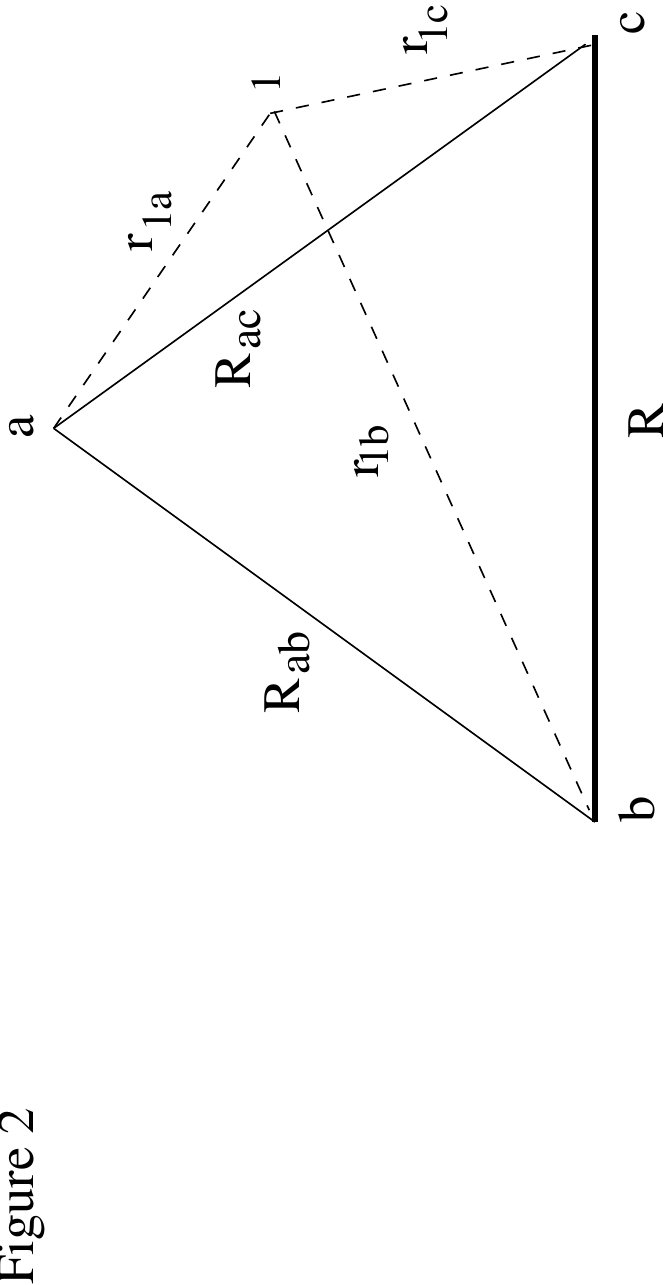}
\end{figure}

\newpage

\textbf{Figure 1:} Definition of the coordinates of one electron with
respect to three centers: linear conformation.

\textbf{Figure 2:} Definition of the coordinates of one electron with
respect to three centers: triangular conformation. 

\end{document}